\documentclass[conference,a4paper]{IEEEtran}
\IEEEoverridecommandlockouts
\usepackage{cite}
\usepackage{amsmath,amssymb,amsfonts}
\usepackage{algorithm}
\usepackage{algorithmic}
\usepackage{graphicx}
\usepackage{textcomp}
\usepackage{xcolor}
\usepackage{url}
\usepackage{eso-pic}
\usepackage{cleveref}
\crefformat{equation}{eq.~(#2#1#3)}
\Crefformat{equation}{eq.~(#2#1#3)}
\crefrangeformat{equation}{eqs.~(#3#1#4)--(#5#2#6)}
\Crefrangeformat{equation}{eqs.~(#3#1#4)--(#5#2#6)}
\crefmultiformat{equation}{eqs.~(#2#1#3)}{ and~(#2#1#3)}{, (#2#1#3)}{ and~(#2#1#3)}
\Crefmultiformat{equation}{eqs.~(#2#1#3)}{ and~(#2#1#3)}{, (#2#1#3)}{ and~(#2#1#3)}
\graphicspath{{figures/}}
\def\BibTeX{{\rm B\kern-.05em{\sc i\kern-.025em b}\kern-.08em
    T\kern-.1667em\lower.7ex\hbox{E}\kern-.125emX}}
\begin{document}

\title{Measurement-Efficient Differentiable Quantum Architecture Search for Combinatorial Optimization}

\author{\IEEEauthorblockN{Lukas Theißinger}
\IEEEauthorblockA{\textit{University of Bonn}\\
Germany\\
lukas.theissinger@uni-bonn.de}
\and
\IEEEauthorblockN{Thore Gerlach}
\IEEEauthorblockA{\textit{European Space Agency}\\
Netherlands}
\and
\IEEEauthorblockN{Christian Bauckhage}
\IEEEauthorblockA{\textit{Lamarr Institute}\\
Germany}
\thanks{This work was funded by the Lamarr Institute in Germany.}
}

\AddToShipoutPictureFG*{%
  \AtPageUpperLeft{%
    \put(\LenToUnit{0.5\paperwidth},\LenToUnit{-0.3cm}){%
      \makebox[0pt][c]{\parbox[t]{0.86\paperwidth}{\centering\scriptsize
        This work has been accepted for publication in the 2026 IEEE 2nd International Conference on Quantum Artificial Intelligence (QAI).\\
        \copyright~2026 IEEE. Personal use of this material is permitted. Permission from IEEE must be obtained for all other uses, in any current or future media, including reprinting/republishing this material for advertising or promotional purposes, creating new collective works, for resale or redistribution to servers or lists, or reuse of any copyrighted component of this work in other works.}}}}}

\maketitle

\begin{abstract}
Differentiable quantum architecture search (DQAS) is a promising framework for the automated design of quantum circuits, particularly for variational quantum optimization algorithms. However, its practical deployment on quantum hardware is limited by the large number of circuit measurements required during optimization, making hardware execution costly.

In this work, we show that for a broad class of combinatorial optimization problems and commonly used rotational gate parameterizations, the measurement cost of DQAS can be significantly reduced without changing the optimization objective. We derive the proposed measurement reduction scheme theoretically and validate it experimentally on 3-SAT and MaxCut benchmark problems. Our approach reduces the requested gradient measurement cost by about 39 to 41\% while introducing only negligible classical post-processing overhead, lowering the practical cost of executing DQAS on quantum hardware.
\end{abstract}


\section{Introduction}
Quantum variational algorithms are a leading approach for using near-term quantum devices on optimization and simulation tasks. In combinatorial optimization, many canonical problems, including 3-SAT, MaxCut, graph coloring, Max-$k$-SAT, graph partitioning and weighted constraint satisfaction, are NP-hard and arise naturally in scheduling, logistics, network design and resource allocation. A key reason these problems are attractive for quantum optimization is that many of them admit compact encodings as Ising or quadratic unconstrained binary optimization (QUBO) Hamiltonians, whose ground states correspond to optimal or feasible high-quality solutions of the original discrete problem \cite{lucas2014ising,glover2019qubo}. These encodings make combinatorial optimization a natural setting for variational quantum algorithms such as the variational quantum eigensolver and the quantum approximate optimization algorithm \cite{peruzzo2014variational,farhi2014quantum,mcclean2016theory,cerezo2021variational}.

A variational quantum algorithm optimizes the parameters of a parameterized quantum circuit so as to minimize the expectation value of a problem Hamiltonian. In the standard setting, the circuit architecture is chosen before training and remains fixed; only the continuous gate parameters are updated. This separation between architecture design and parameter optimization is convenient, but it makes performance highly dependent on a manually selected ansatz. Moreover, gradient-based training on quantum hardware is itself measurement-intensive: analytic parameter-shift rules typically estimate a single rotation-gate derivative from two shifted circuit evaluations \cite{mitarai2018quantum,schuld2019evaluating}.

Differentiable Quantum Architecture Search (DQAS) addresses the fixed-architecture limitation by jointly optimizing continuous gate parameters and discrete circuit choices \cite{zhang2022differentiable}. DQAS introduces a differentiable relaxation of gate selection, allowing architecture weights and gate parameters to be trained together with gradient-based optimization. This places DQAS in a broader line of work that treats ansatz structure as part of the optimization problem, including structure optimization for parameterized quantum circuits, adaptive ansatz construction in ADAPT-VQE and qubit-ADAPT-VQE and differentiable quantum architecture search methods such as QuantumDARTS \cite{ostaszewski2021structure,grimsley2019adaptive,tang2021qubitadapt,wu2023quantumdarts}. These methods differ in how they choose circuit structure: ADAPT-style methods greedily grow the ansatz from an operator pool using gradient information, whereas DQAS-style methods relax discrete architectural choices so that they can be optimized directly.

Despite this flexibility, architecture-adaptive variational algorithms remain expensive to execute on quantum hardware. Each optimization step requires repeated circuit executions to estimate expectation values and gradients, while the associated classical optimization is typically cheap by comparison. Measurement overhead is therefore a central practical bottleneck, especially for methods that evaluate many candidate gates or operators. This issue has been studied explicitly in ADAPT-VQE, where gradient-measurement cost can dominate and specialized measurement strategies are needed to make adaptive ansatz construction practical \cite{anastasiou2023really}.

In this work, we show that the gradient-measurement overhead of DQAS can be reduced substantially for an important class of combinatorial optimization Hamiltonians. We consider Hamiltonians that are diagonal in the computational basis, including Ising and QUBO formulations used for MaxCut, 3-SAT, Max-$k$-SAT, graph partitioning and weighted constraint satisfaction. For DQAS circuits built from parameterized Pauli-string rotation gates, we derive a measurement-saving gradient estimator that replaces a naively applied two-shift parameter-shift estimate with a one-shift estimate whenever the suffix of the circuit has the required diagonal or Clifford structure. The method leaves the optimization objective unchanged: it reduces the number of quantum circuit measurements needed to estimate gradients, while adding only cheap classical post-processing. We prove the correctness of the estimator and evaluate it on 3-SAT and MaxCut benchmarks, where it reduces the requested gradient-measurement cost by approximately 39-41\% with negligible classical overhead.\footnote{Code and data are available at \url{https://github.com/Newida/ME-DQAS}.}

\section{Related Work}

A central limitation of variational quantum algorithms is that their performance depends strongly on the chosen ansatz. Several lines of work therefore treat circuit structure as an optimization variable rather than a fixed design choice. Structure optimization methods optimize the placement and ordering of parameterized gates \cite{ostaszewski2021structure}, while quantum architecture search frameworks search over candidate circuit layouts for variational algorithms \cite{du2022quantumqas,wang2022quantumnas}. Differentiable Quantum Architecture Search (DQAS) makes this search gradient-based by relaxing discrete gate choices into differentiable architecture parameters, so that circuit parameters and architecture weights can be optimized jointly \cite{zhang2022differentiable}. QuantumDARTS follows a similar philosophy for tasks including MaxCut, ground-state estimation and classification \cite{wu2023quantumdarts}. Adaptive ansatz methods such as ADAPT-VQE and qubit-ADAPT-VQE instead grow circuits greedily from operator pools using gradient information \cite{grimsley2019adaptive,tang2021qubitadapt}. Recent work has studied how to reduce the resulting ADAPT-VQE gradient-measurement cost by exploiting Pauli structure and shared measurement information \cite{anastasiou2023really}. Our work targets the corresponding cost inside differentiable architecture search: the shifted circuit evaluations needed to compute gradients for many candidate gates and parameters.

Analytic gradient estimation for variational circuits is commonly based on parameter-shift rules, which express derivatives of expectation values as combinations of expectation values evaluated at shifted parameter values \cite{mitarai2018quantum,schuld2019evaluating}. For gates generated by Pauli strings, the standard rule uses two shifted evaluations per parameter. Later work generalized these rules to broader generator spectra and developed single-component gradient rules for more general variational settings \cite{wierichs2022general,hubregtsen2022singlecomponent}. These results provide general-purpose gradient estimators, but they do not exploit the specific structure that arises in DQAS for diagonal combinatorial Hamiltonians. Our method uses this additional structure: when the cost Hamiltonian is diagonal in the computational basis and the circuit suffix has the required diagonal or Clifford form, one of the two shifted evaluations can be replaced by classical post-processing of a single shifted measurement distribution.

Measurement cost is a major bottleneck in variational quantum algorithms, and many prior approaches reduce this cost by changing how Hamiltonian expectation values are measured. Common strategies include grouping mutually compatible Pauli terms, formulating measurement grouping as a clique-cover problem, using unitary partitioning and constructing deterministic measurement schedules for families of Pauli observables \cite{verteletskyi2020measurement,izmaylov2020unitary,zhao2020measurement,huang2021derandomization}. These techniques are most useful when the Hamiltonian contains many non-commuting Pauli terms. For the diagonal Ising and QUBO Hamiltonians considered in this work, the Hamiltonian terms are already simultaneously measurable in the computational basis. The dominant avoidable cost is therefore not the number of measurement bases for the objective, but the number of distinct shifted circuits required for DQAS gradient estimation. Our contribution targets this different source of overhead by reducing two-shift gradient evaluations to one-shift evaluations under structural conditions that hold for an important class of architecture-search circuits.

\section{Fundamentals}

\subsection{Combinatorial Optimization as a Diagonal Hamiltonian}
\label{subsec:diagonal_hamiltonian}

Many discrete optimization objectives can be encoded as cost Hamiltonians that are diagonal in the computational basis.
This is the standard representation used in Ising and QUBO formulations of combinatorial optimization problems and also underlies QAOA-style cost Hamiltonians \cite{lucas2014ising,glover2019qubo,farhi2014quantum,hadfield2019qaoa}.
Since our measurement-saving rule relies on this diagonal structure, we review the construction explicitly on 3-SAT, a canonical NP-complete constraint satisfaction problem \cite{karp1972reducibility}.

Consider a 3-SAT instance with Boolean variables \(x_1,\ldots,x_n\) and clauses \(c=1,\ldots,m\).
Rather than treating 3-SAT only as a decision problem, we use the equivalent optimization objective that counts violated clauses:
\[
f(x)=\sum_{c=1}^{m} v_c(x),
\qquad
v_c(x)=
\begin{cases}
	1,&\text{if }c\text{ violated by }x,\\
	0,&\text{otherwise}.
\end{cases}
\]
A satisfying assignment exists exactly when \(\min_x f(x)=0\).

The corresponding quantum cost Hamiltonian is defined by assigning each computational basis state the classical objective value, \(H_f\lvert x\rangle = f(x)\lvert x\rangle\). Equivalently, \(H_f=\sum_{x\in\{0,1\}^n} f(x)\lvert x\rangle\langle x\rvert\), which makes the diagonal structure explicit. Of course, this diagonal representation is exponential in the number of qubits.
For 3-SAT, this Hamiltonian can also be written compactly as a sum of clause projectors.
For example, the clause $C=(x_1\vee \neg x_2\vee x_5)$ is violated only when $x_1=0, \text{ } x_2=1,\text{ } x_5=0$.
Its clause Hamiltonian is therefore
\[
H_C
=
\frac{I+Z_1}{2}
\frac{I-Z_2}{2}
\frac{I+Z_5}{2}.
\]
This operator contributes one unit of energy exactly on assignments that violate \(C\) and zero otherwise. Summing over clauses gives \(H_f=\sum_{c=1}^{m} H_c\).
Each \(H_c\) is a product of \(I\) and \(Z\) operators, so all clause terms commute and are simultaneously diagonal in the computational basis.

Given a parameterized quantum circuit \(U(\theta)\), the variational objective is
\begin{equation}
	\label{eq:objective_function}
	F(\theta)
	=
	\langle 0|
	U(\theta)^\dagger
	H_f
	U(\theta)
	|0\rangle .
\end{equation}
If the output state \(U(\theta)\lvert 0\rangle\) is measured in the computational basis, it induces a distribution \(p_\theta\) over bit strings. Because \(H_f\) is diagonal, the objective can be estimated purely from sampled bit strings as \(F(\theta)=\mathbb{E}_{X\sim p_\theta}\left[f(X)\right]\).
In practice, one repeatedly samples \(X\sim p_\theta\), evaluates the classical cost \(f(X)\) and averages the resulting values.

\subsection{Differentiable Quantum Architecture Search (DQAS)}

Differentiable Quantum Architecture Search (DQAS) can be used to minimize the objective in \cref{eq:objective_function}.
It extends variational quantum algorithms by jointly optimizing the circuit architecture and the continuous gate parameters. Instead of optimizing a single fixed circuit, DQAS maintains a parameterized probability distribution \(p_\phi(a)\) over candidate circuit architectures \(a\). During each optimization step, architectures are sampled from this distribution, their circuit parameters are optimized and the resulting objective values are used to update the architecture distribution.

For a sampled architecture \(a\), the circuit parameters \(\theta\) are optimized by minimizing the objective
\[
F_a(\theta)
=
\langle 0|
U_a(\theta)^\dagger
H_f
U_a(\theta)
|0\rangle,
\]
using the parameter-shift rule. For rotation gates generated by Pauli operators, the derivative with respect to a parameter \(\theta_i\) is given by
\[
\frac{\partial F_a(\theta)}{\partial\theta_i}
=
\frac{1}{2}
\left[
F_a\left(\theta+\frac{\pi}{2}e_i\right)
-
F_a\left(\theta-\frac{\pi}{2}e_i\right)
\right],
\]
where \(e_i\) is the unit vector in the \(i\)-th parameter direction.
Each gradient evaluation therefore requires two additional expectation-value estimates beyond the evaluation of the original objective.

After the circuit parameters have been updated, the objective values of the sampled architectures are used to estimate the gradient of the architecture distribution \(p_\phi\) using the REINFORCE estimator. Since this optimization is performed entirely on the classical processor, its computational cost is negligible compared to the repeated execution of quantum circuits.

The dominant cost of DQAS therefore stems from quantum measurements. For every sampled architecture, one measurement-based estimate of the objective is required together with two additional measurements for each parameter-shift gradient evaluation. As the number of sampled architectures and circuit parameters increases, these repeated expectation-value evaluations quickly become the computational bottleneck. Reducing this measurement overhead is therefore essential for making DQAS practical on near-term quantum hardware. In this work, we make a first important step in this direction.

\section{Finding simpler gradients}
\subsection{Pauli-string rotation gradients}

We first consider the common case where the parametrized gate at position \(j\)
is generated by a Pauli string \(V\), with \(R_V(\theta)=\exp\!\left(-i\theta V/2\right)\), \(V^\dagger = V\) and \(V^2 = I\).
Let \(U_{<j}\) denote the part of the circuit before this gate and \(U_{>j}\)
the part after it. We absorb these two parts into the input state and the
observable, respectively: \(|\phi_{<j}\rangle = U_{<j}|0\rangle\) and \(H_{>j} = U_{>j}^{\dagger} H U_{>j}\).
Then the objective can be written locally around the \(j\)-th gate as
\[
F(\theta)
=
\langle \phi_{<j}|
R_V(\theta)^\dagger H_{>j} R_V(\theta)
|\phi_{<j}\rangle .
\]

Differentiating gives
\begin{equation}
	\label{eq:pauli_rotation_gradient}
	F'(\theta)
	=
	\left\langle \phi_{<j} \right|
	R_V(\theta)^\dagger
	\frac{i}{2}[V,H_{>j}]
	R_V(\theta)
	\left|\phi_{<j}\right\rangle .
\end{equation}
Equivalently, if we define \(|\phi_j(\theta)\rangle = R_V(\theta)|\phi_{<j}\rangle\), then
\begin{equation}
	\label{eq:pauli_rotation_gradient_state}
	F'(\theta)
	=
	\frac{i}{2}
	\langle \phi_j(\theta)|
	[V,H_{>j}]
	|\phi_j(\theta)\rangle .
\end{equation}

The exact zero-gradient condition follows from \cref{eq:pauli_rotation_gradient_state}: \([V,H_{>j}] = 0\).
When this holds, \(F'(\theta)=0\) and no measurement is required for this
parameter. Although checking this condition may require constructing the
dressed observable \(H_{>j}\), there are two cheap sufficient tests.

First, if \(V\) commutes with the original observable \(H\) and with every gate
after position \(j\), then \(V\) also commutes with \(H_{>j}\).
Second, if \(V\) also commutes with \(H\) and every
gate after position \(j\) commutes with \(H\), then again \([V,H_{>j}] = 0\);
therefore, the gradient is again exactly zero.

These conditions are especially cheap to check for the \(Z\)-only Hamiltonians
that arise in many combinatorial optimization problems. In this setting, the
Pauli decomposition of \(H\) is given directly by the problem instance, so we
do not need to construct or measure any new observable. Checking whether \(V\)
commutes with \(H\) only requires checking whether \(V\) commutes with each
Pauli term already present in this known decomposition. Similarly, checking
whether a later Pauli-rotation gate commutes with \(H\) only requires checking
commutation between its generator and these same known Pauli terms. Thus both
tests are purely classical preprocessing steps and add negligible overhead
compared with estimating the gradient by circuit measurements.

More generally, suppose that the dressed observable has a Pauli decomposition
\[
H_{>j} = \sum_{\ell} h_\ell P_\ell,
\qquad
h_\ell \in \mathbb{R},
\qquad
P_\ell^\dagger=P_\ell,
\qquad
P_\ell^2=I .
\]
Substituting this into \cref{eq:pauli_rotation_gradient_state} yields
\[
F'(\theta)
=
\frac{1}{2}
\sum_\ell h_\ell
\langle \phi_j(\theta)|
i[V,P_\ell]
|\phi_j(\theta)\rangle .
\]
Since both \(V\) and \(P_\ell\) are Pauli strings, they either commute or
anticommute. If \([V,P_\ell]=0\), then the corresponding term contributes
nothing to the gradient. If \(VP_\ell=-P_\ell V\), then \([V,P_\ell] = 2VP_\ell\).
Therefore only Pauli terms that anticommute with \(V\) contribute:
\begin{equation}
	\label{eq:anti_objective}
	F'(\theta)
	=
	\sum_{\ell\in\mathcal A(V)}
	h_\ell
	\langle \phi_j(\theta)|
	iVP_\ell
	|\phi_j(\theta)\rangle ,
\end{equation}
where \(\mathcal A(V)=\{\ell : VP_\ell=-P_\ell V\}\).

For every \(\ell\in\mathcal A(V)\), we have the identity
\(R_V(\pi/2)^\dagger P_\ell R_V(\pi/2)=iVP_\ell\).
Therefore,
\[
\begin{aligned}
	F'(\theta)
	&=
	\sum_{\ell\in\mathcal A(V)}
	h_\ell
	\langle \phi_j(\theta)|
	R_V(\pi/2)^\dagger P_\ell R_V(\pi/2)
	|\phi_j(\theta)\rangle \\
	&=
	\sum_{\ell\in\mathcal A(V)}
	h_\ell
	\langle \phi_{<j}|
	R_V(\theta+\pi/2)^\dagger
	P_\ell
	R_V(\theta+\pi/2)
	|\phi_{<j}\rangle .
\end{aligned}
\]
Equivalently, if we define the anticommuting part of the Hamiltonian as \(H_{\mathcal A(V)}=\sum_{\ell\in\mathcal A(V)} h_\ell P_\ell\), then
\begin{equation}
	\label{eq:one_shift_gradient}
	F'(\theta)
	=
	\langle \phi_j(\theta + \pi/2)|
	H_{\mathcal A(V)}
	|\phi_j(\theta + \pi/2)\rangle .
\end{equation}

Thus the gradient can be estimated by measuring only these anticommuting terms and shifting the parameter by $\pi/2$.

\subsection{Cheaper gradients for \(Z\)-only Hamiltonians}

For combinatorial optimization Hamiltonians that are diagonal in the computational basis, the gradient can often be measured at substantially lower cost. In the 3-SAT case, we have seen that the problem Hamiltonian consists only of \(Z\)-type Pauli strings $H=\sum_{P_\ell} h_\ell P_\ell, \text{ } P_\ell \in\{I,Z\}^{\otimes n}$.
The same structure appears in many other combinatorial optimization problems, including MaxCut, QUBO, Ising models and graph coloring. In these settings, the Pauli decomposition of $H$ is not an additional object that must be constructed or measured term by term from scratch; it is already specified by the problem instance.

\subsubsection{Commuting suffix}

Starting again from \cref{eq:pauli_rotation_gradient_state}, assume that all gates applied after position \(j\) commute with the original
Hamiltonian \(H\). Then \(H_{>j}=U_{>j}^{\dagger} H U_{>j}=H\).
Thus the gradient from \cref{eq:one_shift_gradient} simplifies. It no longer requires the dressed Hamiltonian $H_{>j}$, but now consists of the original problem Hamiltonian.
Since $H$ is diagonal, \(H_{\mathcal A(V)}\) is also \(Z\)-diagonal and it can be measured in the computational basis, just like the original combinatorial optimization Hamiltonian.
Thus the gradient can be calculated by sampling $X \sim p_{\theta + \pi/2}$ and averaging the diagonal entries \(f_{\mathcal A(V)}(X)\) of \(H_{\mathcal A(V)}\).
This has no additional costs, since no gate type or non-standard measurement basis is required.

Compared with the standard parameter-shift rule, \(F'(\theta)=\frac{1}{2}\left(F(\theta+\pi/2)-F(\theta-\pi/2)\right)\), this method requires only one shifted circuit evaluation. Moreover, the
measurement only needs the Hamiltonian terms that anticommute with \(V\); the
commuting terms are known classically to have zero contribution to the
gradient.

\subsubsection{Clifford suffix}

A more general useful case occurs when the suffix after position \(j\) does	
not commute with \(H\), but is a Clifford circuit. Let the suffix be
\(C_{>j}\). The objective function becomes \(\langle C_{>j} \phi_j(\theta)| H | C_{>j} \phi_j(\theta) \rangle\).
Therefore, the gradient is
$$
F'(\theta)
=
\langle C_{>j}\phi_j(\theta + \pi/2)|
H_{\mathcal A_C(V)}
|C_{>j}\phi_j(\theta + \pi/2)\rangle ,
$$
where \(H_{\mathcal A_C(V)}\) keeps the terms \(h_\ell P_\ell\) of \(H\) for which \(C_{>j}^\dagger P_\ell C_{>j}\) anticommutes with \(V\).

Hence, the gradient can be calculated by measuring the circuit $C_{>j} |\phi_j(\theta + \pi/2)\rangle$.
Compared to two of the naive parameter-shift rule this requires only one measurement in the computational basis.
On the other hand, this introduces a quantum cost, because even though the quantum circuit technically stays the same length,
because the clifford part was a suffix of the circuit it is usually handled classically as post-processing.
This can no longer be done if this clifford part is introduced at the beginning of the circuit.
However, it does not introduce any new measurement groups, which a classical handling of the clifford part could add additionally.
In summary, if the transformed observable does not remain measurable in one cheap group and if the Clifford suffix is shallow, then physically applying the suffix may be better, than trying to handle it classically only.

\begin{figure*}[t]
\centering
\includegraphics[width=0.94\textwidth]{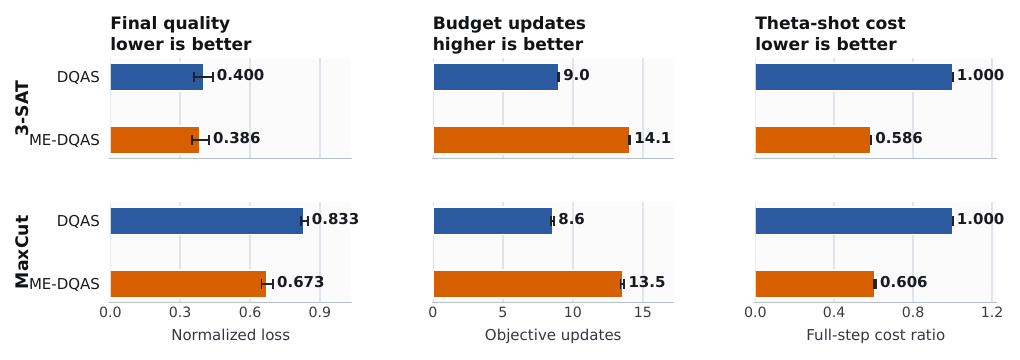}
\caption{Main experimental result for 100 matched 3-SAT instances and 100
matched MaxCut instances. Each row compares baseline DQAS with ME-DQAS under
the same global shot budget. Bars show means with 95\% confidence intervals.
The three panels measure final solution quality, how many DQAS updates fit into
the budget and the requested full \(\theta\)-gradient shot cost relative to
baseline parameter shift.}
\label{fig:bar-summary}
\end{figure*}

\begin{figure}[t]
\centering
\includegraphics[width=\linewidth]{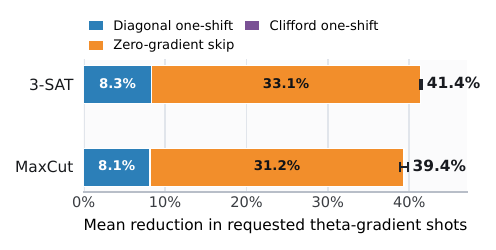}
\caption{Source of the requested full \(\theta\)-shot reduction in
\cref{fig:bar-summary}. The dominant contribution comes from gradients that are
certified to be exactly zero; the diagonal one-shift rule gives the next largest
contribution. The Clifford-suffix branch is negligible for this gate set.}
\label{fig:reduction-breakdown}
\end{figure}

\subsection{Measurement-Efficient DQAS}

The gradient identities above lead to a direct modification of DQAS. The
architecture-sampling procedure, objective evaluation and architecture update
are kept unchanged. ME-DQAS differs from ordinary DQAS only in how it estimates
gradients with respect to circuit parameters \(\theta\).

For each parameterized Pauli rotation in a sampled architecture, the algorithm
first checks whether the derivative can be certified to vanish by commutation.
If not, it checks whether the suffix structure allows the gradient to be
estimated from one shifted circuit, either because the suffix preserves the
diagonal Hamiltonian or because it is Clifford. Only gates that satisfy none of
these conditions use the ordinary two-shift parameter-shift estimator.

The resulting procedure is shown in \cref{alg:me-dqas}. The highlighted lines
are the algorithmic changes relative to baseline DQAS.

\begin{algorithm}[ht]
	\caption{ME-DQAS}
	\label{alg:me-dqas}
	\begin{algorithmic}[1]
		\REQUIRE \(L\) circuit components; operator pool \(\mathcal O\); model \(P_\phi\); batch size \(K\); Hamiltonian \(H_f\)
		\WHILE{not converged}
		\STATE Sample a batch of \(K\) circuits from \(P_\phi\)
		\STATE Compute the objective \cref{eq:objective_function} for each circuit in the batch
		
		\FOR{each parameterized gate \(R_V(\theta_j)\) in the sampled circuits}
		\STATE \textcolor{blue}{Let \(U_{>j}\) denote the part of the circuit after \(R_V(\theta_j)\)}
		\STATE \textcolor{blue}{Compute available commutation certificates:}
		\STATE \textcolor{blue}{$c_{VH} := [V,H_f]=0$}
		\STATE \textcolor{blue}{$c_{VU} := [V,U_{>j}]=0$}
		\STATE \textcolor{blue}{$c_{HU} := [H_f,U_{>j}]=0$}
		
		\IF{\textcolor{blue}{\(c_{VH}\) and \((c_{VU}\) or \(c_{HU})\)}}
		\STATE \textcolor{blue}{Set \(\nabla_{\theta_j} f = 0\)}
		\ELSIF{\textcolor{blue}{\(c_{HU}\)}}
		\STATE \textcolor{blue}{Estimate \(\nabla_{\theta_j} f\) from one shifted circuit at \(\theta_j+\pi/2\)}
		\ELSIF{\textcolor{blue}{\(U_{>j}\) is Clifford}}
		\STATE \textcolor{blue}{Estimate \(\nabla_{\theta_j} f\) using the circuit $C_{>j}|\phi_j(\theta + \pi/2)\rangle$}
		\ELSE
		\STATE Estimate \(\nabla_{\theta_j} f\) using the ordinary two-shift parameter-shift estimator
		\ENDIF
		\ENDFOR
		
		\STATE Compute the gradients with respect to \(\phi\)
		\STATE Update \(\theta\) and \(\phi\) with gradient-based optimizers
		\ENDWHILE
	\RETURN circuit found with the lowest objective evaluation
\end{algorithmic}
\end{algorithm}

The checks in \cref{alg:me-dqas} are purely structural. They use the known
Pauli decomposition of the diagonal problem Hamiltonian and the generators of
the gates appearing after the differentiated parameter. Thus they add only
a small overhead and do not change the objective being optimized.

The first branch implements the zero-gradient certificates from
\cref{eq:pauli_rotation_gradient_state}. The second branch applies the
one-shift rule from \cref{eq:one_shift_gradient} when the circuit suffix
commutes with \(H_f\), so that only the anticommuting diagonal terms contribute
to the derivative. The third branch handles the Clifford-suffix case, where the
suffix can be propagated through Pauli observables while preserving Pauli
structure. In all remaining cases, ME-DQAS falls back to the standard
two-shift parameter-shift rule.

Consequently, ME-DQAS preserves the DQAS search objective and update structure.
Its effect is limited to reducing the number of shifted circuit evaluations
requested for eligible \(\theta\)-gradient components. This is the measurement
cost reported in the experiments below.

The probability that a gradient falls into one of the cheaper branches is
problem- and architecture-dependent. It increases when the Hamiltonian is
sparse or diagonal, when the gate pool contains many diagonal, commuting or
Clifford operations and when the sampled suffix after the differentiated gate
preserves this structure. It decreases when dense Hamiltonians or non-Clifford
mixing gates make commutation with the suffix unlikely. Circuit depth therefore
has two competing effects: longer suffixes create more chances to break the
required structure, but a searched architecture may also learn suffixes rich in
commuting or Clifford gates. We measure these effects empirically in the
breakdown experiment below.

\section{Experiments}

\subsection{Protocol and Metrics}

We evaluate ME-DQAS against baseline DQAS using paired, shot-budgeted
simulations. In each pair, both algorithms receive the same problem instance,
initial \(\theta\), architecture-distribution seed, batch size, learning rates
and global shot budget. The main difference is the estimator used for
\(\theta\)-gradients: baseline DQAS applies the two-shift parameter-shift rule
to every sampled parametric gate, while ME-DQAS uses the zero-gradient,
one-shift diagonal, one-shift Clifford or fallback branch from
\cref{alg:me-dqas}.

The benchmarks use two diagonal objectives. For 3-SAT, the loss is the number
of violated clauses. For MaxCut, the problem is written as a minimization task
by counting uncut edges. In both benchmarks, \(n=8\), the circuit length is
\(L=8\) and each architecture position chooses from
\(\{R_x,R_y,R_z,\mathrm{CZ}\}\). Random 3-SAT formulas and random MaxCut graphs
use \(m=\alpha n\) local objective terms with \(\alpha\in\{1.5,2.0\}\). All
runs use global budget \(B=750{,}000\), batch size \(K=128\),
\(S_f=S_\theta=50\), one \(\theta\)-update per DQAS iteration, architecture
learning rate \(0.3\), \(\theta\)-learning rate \(0.15\), initial \(\theta\)
scale \(0.4\) and a cap of 100 iterations.

\Cref{fig:bar-summary} reports final normalized best-so-far loss, the number of
completed objective updates and the requested full \(\theta\)-gradient
shot-cost ratio. For quality, each run is normalized by the exact objective
range,
\[
\widetilde F_t=
\frac{\min_{\tau\le t}F_\tau-F_{\min}^{\mathrm{exact}}}
{F_{\max}^{\mathrm{exact}}-F_{\min}^{\mathrm{exact}}}.
\]
ME-DQAS improves all three metrics. It reduces the requested
\(\theta\)-gradient shot cost to \(0.5857\pm0.0014\) of baseline on 3-SAT and
\(0.6064\pm0.0046\) on MaxCut, corresponding to \(41.4\%\) and \(39.4\%\)
fewer requested gradient shots. Under the same global shot budget, this buys
roughly five additional DQAS updates: \(14.05\pm0.04\) instead of
\(8.99\pm0.02\) on 3-SAT and \(13.54\pm0.13\) instead of \(8.56\pm0.10\) on
MaxCut. The additional search progress coincides with lower final normalized
loss on both benchmarks.

\Cref{fig:reduction-breakdown} shows that most savings come from complete
gradient removal, not from a uniform constant-factor reduction. The
zero-gradient certificate saves \(33.1\%\) of baseline \(\theta\)-shot cost on
3-SAT and \(31.2\%\) on MaxCut; the diagonal one-shift rule adds \(8.3\%\) and
\(8.1\%\). This matches the structural discussion above. With a \(Z\)-diagonal
Hamiltonian and gate set \(\{R_x,R_y,R_z,\mathrm{CZ}\}\), many suffixes contain
diagonal or diagonal-preserving operations, while the Clifford branch is rarely
available for arbitrary-angle rotations. ME-DQAS therefore changes the
measurement budget, not the objective landscape: it evaluates the same
objective and uses the same outer DQAS update rule, but measures eligible
\(\theta\)-gradients more cheaply.

\section{Conclusion}

We introduced ME-DQAS, a measurement-efficient variant of differentiable quantum
architecture search for diagonal combinatorial Hamiltonians. The method uses
commutation certificates and one-shift rules to reduce the number of circuit
evaluations requested for eligible \(\theta\)-gradients while leaving the DQAS
objective, architecture sampling and outer update rule unchanged.

On paired 3-SAT and MaxCut benchmarks, ME-DQAS reduced the requested full
\(\theta\)-gradient shot cost to \(58.6\%\) and \(60.6\%\) of baseline DQAS.
Under the same global shot budget, this produced roughly five additional
architecture-search updates and better final normalized best-so-far loss. The
fraction of eligible gradients will depend on the Hamiltonian, gate pool,
suffix structure and hardware noise, but the mechanism is general: when the
cost Hamiltonian and circuit suffix expose commutation or Clifford structure,
DQAS gradients need not be measured with the full two-shift rule.

\bibliographystyle{IEEEtran}
\bibliography{references}

\begin{thebibliography}{10}
\providecommand{\url}[1]{#1}
\csname url@samestyle\endcsname
\providecommand{\newblock}{\relax}
\providecommand{\bibinfo}[2]{#2}
\providecommand{\BIBentrySTDinterwordspacing}{\spaceskip=0pt\relax}
\providecommand{\BIBentryALTinterwordstretchfactor}{4}
\providecommand{\BIBentryALTinterwordspacing}{\spaceskip=\fontdimen2\font plus
\BIBentryALTinterwordstretchfactor\fontdimen3\font minus \fontdimen4\font\relax}
\providecommand{\BIBforeignlanguage}[2]{{%
\expandafter\ifx\csname l@#1\endcsname\relax
\typeout{** WARNING: IEEEtran.bst: No hyphenation pattern has been}%
\typeout{** loaded for the language `#1'. Using the pattern for}%
\typeout{** the default language instead.}%
\else
\language=\csname l@#1\endcsname
\fi
#2}}
\providecommand{\BIBdecl}{\relax}
\BIBdecl

\bibitem{lucas2014ising}
A.~Lucas, ``{Ising} formulations of many {NP} problems,'' \emph{Frontiers in Physics}, vol.~2, p.~5, 2014.

\bibitem{glover2019qubo}
F.~Glover, G.~Kochenberger, and Y.~Du, ``Quantum bridge analytics {I}: A tutorial on formulating and using {QUBO} models,'' \emph{4OR}, vol.~17, no.~4, pp. 335--371, 2019.

\bibitem{peruzzo2014variational}
A.~Peruzzo, J.~McClean, P.~Shadbolt, M.-H. Yung, X.-Q. Zhou, P.~J. Love, A.~Aspuru-Guzik, and J.~L. O'Brien, ``A variational eigenvalue solver on a photonic quantum processor,'' \emph{Nature Communications}, vol.~5, no.~1, p. 4213, 2014.

\bibitem{farhi2014quantum}
\BIBentryALTinterwordspacing
E.~Farhi, J.~Goldstone, and S.~Gutmann, ``A quantum approximate optimization algorithm,'' 2014, arXiv:1411.4028. [Online]. Available: \url{https://arxiv.org/abs/1411.4028}
\BIBentrySTDinterwordspacing

\bibitem{mcclean2016theory}
J.~R. McClean, J.~Romero, R.~Babbush, and A.~Aspuru-Guzik, ``The theory of variational hybrid quantum-classical algorithms,'' \emph{New Journal of Physics}, vol.~18, no.~2, p. 023023, 2016.

\bibitem{cerezo2021variational}
M.~Cerezo, A.~Arrasmith, R.~Babbush, S.~C. Benjamin, S.~Endo, K.~Fujii, J.~R. McClean, K.~Mitarai, X.~Yuan, L.~Cincio, and P.~J. Coles, ``Variational quantum algorithms,'' \emph{Nature Reviews Physics}, vol.~3, no.~9, pp. 625--644, 2021.

\bibitem{mitarai2018quantum}
K.~Mitarai, M.~Negoro, M.~Kitagawa, and K.~Fujii, ``Quantum circuit learning,'' \emph{Physical Review A}, vol.~98, no.~3, p. 032309, 2018.

\bibitem{schuld2019evaluating}
M.~Schuld, V.~Bergholm, C.~Gogolin, J.~Izaac, and N.~Killoran, ``Evaluating analytic gradients on quantum hardware,'' \emph{Physical Review A}, vol.~99, no.~3, p. 032331, 2019.

\bibitem{zhang2022differentiable}
S.-X. Zhang, C.-Y. Hsieh, S.~Zhang, and H.~Yao, ``Differentiable quantum architecture search,'' \emph{Quantum Science and Technology}, vol.~7, no.~4, p. 045023, 2022.

\bibitem{ostaszewski2021structure}
M.~Ostaszewski, E.~Grant, and M.~Benedetti, ``Structure optimization for parameterized quantum circuits,'' \emph{Quantum}, vol.~5, p. 391, 2021.

\bibitem{grimsley2019adaptive}
H.~R. Grimsley, S.~E. Economou, E.~Barnes, and N.~J. Mayhall, ``An adaptive variational algorithm for exact molecular simulations on a quantum computer,'' \emph{Nature Communications}, vol.~10, no.~1, p. 3007, 2019.

\bibitem{tang2021qubitadapt}
H.~L. Tang, V.~O. Shkolnikov, G.~S. Barron, H.~R. Grimsley, N.~J. Mayhall, E.~Barnes, and S.~E. Economou, ``{Qubit}-{ADAPT}-{VQE}: An adaptive algorithm for constructing hardware-efficient ans{\"a}tze on a quantum processor,'' \emph{PRX Quantum}, vol.~2, no.~2, p. 020310, 2021.

\bibitem{wu2023quantumdarts}
\BIBentryALTinterwordspacing
W.~Wu, G.~Yan, X.~Lu, K.~Pan, and J.~Yan, ``{QuantumDARTS}: Differentiable quantum architecture search for variational quantum algorithms,'' in \emph{Proceedings of the 40th International Conference on Machine Learning}, ser. Proceedings of Machine Learning Research, vol. 202.\hskip 1em plus 0.5em minus 0.4em\relax PMLR, 2023, pp. 37\,745--37\,764. [Online]. Available: \url{https://proceedings.mlr.press/v202/wu23v.html}
\BIBentrySTDinterwordspacing

\bibitem{anastasiou2023really}
\BIBentryALTinterwordspacing
P.~G. Anastasiou, N.~J. Mayhall, E.~Barnes, and S.~E. Economou, ``How to really measure operator gradients in {ADAPT}-{VQE},'' 2023, arXiv:2306.03227. [Online]. Available: \url{https://arxiv.org/abs/2306.03227}
\BIBentrySTDinterwordspacing

\bibitem{du2022quantumqas}
Y.~Du, T.~Huang, S.~You, M.-H. Hsieh, and D.~Tao, ``Quantum circuit architecture search for variational quantum algorithms,'' \emph{npj Quantum Information}, vol.~8, no.~1, p.~62, 2022.

\bibitem{wang2022quantumnas}
H.~Wang, Y.~Ding, J.~Gu, Y.~Lin, D.~Z. Pan, F.~T. Chong, and S.~Han, ``{QuantumNAS}: Noise-adaptive search for robust quantum circuits,'' in \emph{2022 IEEE International Symposium on High-Performance Computer Architecture (HPCA)}, 2022, pp. 692--708.

\bibitem{wierichs2022general}
D.~Wierichs, J.~Izaac, C.~Wang, and C.~Y.-Y. Lin, ``General parameter-shift rules for quantum gradients,'' \emph{Quantum}, vol.~6, p. 677, 2022.

\bibitem{hubregtsen2022singlecomponent}
T.~Hubregtsen, F.~Wilde, S.~Qasim, and J.~Eisert, ``Single-component gradient rules for variational quantum algorithms,'' \emph{Quantum Science and Technology}, vol.~7, no.~3, p. 035008, 2022.

\bibitem{verteletskyi2020measurement}
V.~Verteletskyi, T.-C. Yen, and A.~F. Izmaylov, ``Measurement optimization in the variational quantum eigensolver using a minimum clique cover,'' \emph{The Journal of Chemical Physics}, vol. 152, no.~12, p. 124114, 2020.

\bibitem{izmaylov2020unitary}
A.~F. Izmaylov, T.-C. Yen, R.~A. Lang, and V.~Verteletskyi, ``Unitary partitioning approach to the measurement problem in the variational quantum eigensolver method,'' \emph{Journal of Chemical Theory and Computation}, vol.~16, no.~1, pp. 190--195, 2020.

\bibitem{zhao2020measurement}
A.~Zhao, A.~Tranter, W.~M. Kirby, S.~F. Ung, A.~Miyake, and P.~J. Love, ``Measurement reduction in variational quantum algorithms,'' \emph{Physical Review A}, vol. 101, no.~6, p. 062322, 2020.

\bibitem{huang2021derandomization}
H.-Y. Huang, R.~Kueng, and J.~Preskill, ``Efficient estimation of pauli observables by derandomization,'' \emph{Physical Review Letters}, vol. 127, no.~3, p. 030503, 2021.

\bibitem{hadfield2019qaoa}
S.~Hadfield, Z.~Wang, B.~O'Gorman, E.~G. Rieffel, D.~Venturelli, and R.~Biswas, ``From the quantum approximate optimization algorithm to a quantum alternating operator ansatz,'' \emph{Algorithms}, vol.~12, no.~2, p.~34, 2019.

\bibitem{karp1972reducibility}
R.~M. Karp, ``Reducibility among combinatorial problems,'' in \emph{Complexity of Computer Computations}, R.~E. Miller, J.~W. Thatcher, and J.~D. Bohlinger, Eds.\hskip 1em plus 0.5em minus 0.4em\relax New York: Plenum Press, 1972, pp. 85--103.

\end{thebibliography}

\end{document}